\documentclass[reqno]{amsart}
\usepackage{amsmath}
\usepackage{amssymb}
\usepackage{amsfonts}
\usepackage{mathrsfs}
\usepackage[all,arc]{xy}
\usepackage{enumerate}
\usepackage{mathrsfs}
\usepackage{amsthm}
\usepackage{enumitem}
\usepackage{extarrows}
\usepackage{verbatim}
\usepackage{dsfont}
\usepackage{xcolor}
\usepackage{textcomp}
\usepackage{tikz-cd}
\usepackage{smartdiagram}
\usepackage{tensor}
\usepackage{mathtools}
\usepackage{float}
\usepackage[toc,page]{appendix}
\usepackage{enumitem}
\usetikzlibrary{decorations.pathmorphing}
\usetikzlibrary{backgrounds, chains, positioning, shadows}

\theoremstyle{definition}

\newtheorem{rem}{Remark}
\newtheorem{stp}{Step}

\makeatletter

\newcommand*{\rom}[1]{\expandafter\@slowromancap\romannumeral #1@}

\makeatletter
\begin{document} 
	\title{A precise machine learning aided algorithm for land subsidence or upheave prediction from GNSS time series}   
	\author[M. Kiani]{M. Kiani$^\dag$\\
		$^\dag$ \tiny Graduated from School of surveying and geospatial data engineering, University of Tehran, Tehran, Iran}
	
	\thanks{Corresponding author, email: mostafakiani@alumni.ut.ac.ir, tel:+989100035865}
	\date{}
\begin{abstract}
This paper is aimed at the problem of predicting the land subsidence or upheave in an area, using GNSS position time series. Since machine learning algorithms have presented themselves as strong prediction tools in different fields of science, we employ them to predict the next values of the GNSS position time series. For this reason, we present an algorithm that takes advantage of the machine learning algorithms for the prediction of positions in a GNSS time series. The proposed algorithm has two steps-preprocessing and prediction. In the preprocessing phase, the periodic tidal and atmospheric signals in the time series are removed and coordinates are transferred to the local coordinate system. In the prediction phase, eight different machine learning algorithms are used, namely, multilayer perceptron, Bayesian neural network, radial basis functions, Gaussian processes, k-nearest neighbor, generalized regression neural network, classification and regression trees, and support vector regression. We show the superiority of the Gaussian processes algorithm, compared to other methods, in 14 different real GNSS time series studies. The proposed algorithm can achieve up to 4 millimeters in accuracy, with the average accuracy as 2 centimeters across all time series.        
\end{abstract}
\maketitle
$Key words:$ GNSS position time series, machine learning, training data, prediction accuracy, land subsidence or upheave
\section{Introduction}
Land subsidence or upheave problem in an area is of foremost importance for many applications. Various authors, including \cite{Hu} and \cite{Strozzi}, have developed different methods to deal with this problem. The mentioned papers have assessed the land subsidence or upheave after their occurrence, meaning the evaluation of the land movement is performed based on the assumption that the region has a particular trend. This is, of course, important to possibly avoid the similar trends in future. 

It is interesting to investigate methods that can possibly predict the land movement, either subsidence or upheave. One way to do this is to use the traditional statistical methods like Theta \cite{Assimakopoulos}. However, experiences with these methods \cite{Kiani1} have shown that they are not as strong as machine learning methods, schemes the usage of which enable us to predict the next outcomes of the time series with higher accuracy, speed, and efficiency \cite{Ahmed}, \cite{Makridakis}. Based on this, we have devised a new method for this special problem. Besides using the (conventional) machine learning methods, the precise setting of the problem, including transformations for converting the coordinates to the local coordinate system are used. It is important to notice that, as also \cite{Kiani2} asserts, the machine learning algorithms we use are based on the supervised learning for prediction, which gives the idea of extrapolation, like the numerical methods in \cite{Kiani2}, in contrast to the most methods of approximation in geosciences \cite{Kiani3}-\cite{Kiani7} and \cite{Kiani8}-\cite{Kiani10}. 

The rest of this paper is organized as follows. In section 2 the algorithm is explained. In section 3, a study is presented for the evaluation of method for the time series prediction. Finally, conclusions are mentioned in section 4.     
\section{Explanation of the algorithm}
In this section, the algorithm for land subsidence or upheave prediction is explained in detail. The algorithm to be proposed is based on two distinct steps: preprocessing and prediction. 

\subsection{Preprocessing phase}
The land subsidence or upheave prediction problem from GNSS position time series deals with the concept of trends, i.e. ascending or descending changing behavior of the time series. Therefore, periodic signals should be removed to achieve a better performance for the prediction. On the other hand, periodic atmospheric and tidal signals tend to impact the positions \cite{Bogusz}. Thus, these effects must be removed to achieve a more realistic prediction for subsidence or upheave. In order to remove these untoward contributions, we use the simple parametric model, with the coefficients determined from the time series itself. 

In the following $y_k$ represents the value of time series at time $t_k$ (different $k$ values correspond to different times). In order to find the periodic signals more accurately, first the least squares line for trends is determined and subsequently subtracted from the values $y_k$ to get the periodic values, denoted by $y_k^p$. In mathematical representation, we denote the least squares line by $y_k^{l}=ct_k+d$, the coefficients of which are calculated as the following, based on the $n$ training data

\begin{equation}\label{eqn1}
\begin{bmatrix}
y_1\\
y_2\\
\vdots\\
y_n
\end{bmatrix}
=
\begin{bmatrix}
1 & t_1\\
1 & t_2\\
\vdots\\
1 & t_n
\end{bmatrix}
\begin{bmatrix}
c\\
d
\end{bmatrix}
\end{equation}
\begin{equation}\label{eqn4}
y_k^p=y_k-y_k^l.
\end{equation}
The periodic, parametric model of degree $m$ for atmospheric and tidal effects removal is as follows.
\begin{equation}\label{eqn2}
y_k^p=\sum_{i=1}^{m}a_i\cos(2\pi f_i t_k)+b_i\sin(2\pi f_i t_k),
\end{equation}       
where $f_i~i=1,...,m$ are the frequencies of the mentioned signals, and $a_i,~b_i~i=1,...,m$ are the coefficients that are determined based on the $n$ training data, as the following
\begin{equation}\label{eqn3}
\begin{bmatrix}
y_1^p\\
y_2^p\\
\vdots\\
y_n^p
\end{bmatrix}
=
\begin{bmatrix}
\cos(2\pi f_1 t_1) & \sin(2\pi f_1 t_1) & \cos(2\pi f_2 t_1) & \sin(2\pi f_2 t_1) & \cdot & \cos(2\pi f_m t_1) & \sin(2\pi f_m t_1)\\
\cos(2\pi f_1 t_2) & \sin(2\pi f_1 t_2) & \cos(2\pi f_2 t_2) & \sin(2\pi f_2 t_2) & \cdot & \cos(2\pi f_m t_2) & \sin(2\pi f_m t_2)\\
\vdots\\
\cos(2\pi f_1 t_n) & \sin(2\pi f_1 t_n) & \cos(2\pi f_2 t_n) & \sin(2\pi f_2 t_n) & \cdot & \cos(2\pi f_m t_n) & \sin(2\pi f_m t_n)
\end{bmatrix}
\begin{bmatrix}
a_1\\
b_1\\
a_2\\
b_2\\
\vdots\\
a_m\\
b_m
\end{bmatrix}
\end{equation}
The choice of the frequencies will affect the prediction accuracy in the prediction phase. The most important frequencies of the atmospheric and tidal signals are given in Table \ref{tab1} (refer, for instance, to \cite{Jin} for more information).
\begin{table}[H]
	\centering
	\caption{The most important frequencies of the atmospheric and tidal signals}
	\label{tab1}
	\begin{tabular}{|c|c|}\hline
		\cline{1-2} frequency & frequency value ($(siderial~year)^{-1}$)\\
		
        \cline{1-2} $f_1$ & 1.3689 $\times 10^{-3}$\\
        
        \cline{1-2} $f_2$ & 2.7378 $\times 10^{-3}$\\
        
        \cline{1-2} $f_3$ & 3.8329 $\times 10^{-2}$\\
        
        \cline{1-2} $f_4$ & 8.2134 $\times 10^{-2}$\\
        
        \cline{1-2} $f_5$ & 4.9308 $\times 10^{-1}$\\
        
        \cline{1-2} $f_6$ & 9.8424 $\times 10^{-1}$\\

		\hline
	\end{tabular}
\end{table}
\begin{rem}\label{rem1}
In fact the mentioned frequencies in Table \ref{tab1} are the inverse of fundamental atmospheric and tidal periods, which are 0.5, 1, 14, 30, 180.1, and 359.5 days.  
\end{rem}
After calculating the periodic signals contribution, these values must be subtracted from the original data as follows, to get the trend-only time series values-$y_k^T$. 
\begin{equation}\label{eqn5}
y_k^T=y_k-y_k^p.
\end{equation}
After this step, the derived values are transferred to the local coordinate system, since the subsidence or upheave would be more tangibly understood. The steps are as follows.
\begin{stp}\label{stp1}
The geodetic coordinates, $(\phi,\lambda,h)$, of the points in the time series, on the WGS84 ellipsoid with the specific semi-major axis, $a$, and eccentricity, $e^2$ values, are calculated based on the three dimensional $(X,Y,Z)$ values, in an iterative manner \cite{Jekeli}, \cite{Deakin}
\begin{equation}\label{eqn6}
\tan\lambda=\frac{Y}{X},
\end{equation}
\begin{equation}\label{eqn7}
\begin{split}
P&=\sqrt{X^2+Y^2},\\
R_N^0&=a,\\
h_0&=\sqrt{X^2+Y^2+Z^2}-a\sqrt[4]{1-e^2},\\
\phi_0&=\arctan(\frac{Z}{P}(1-e^2\frac{N_0}{N_0+h_0})^{-1}),
\end{split}
\end{equation}
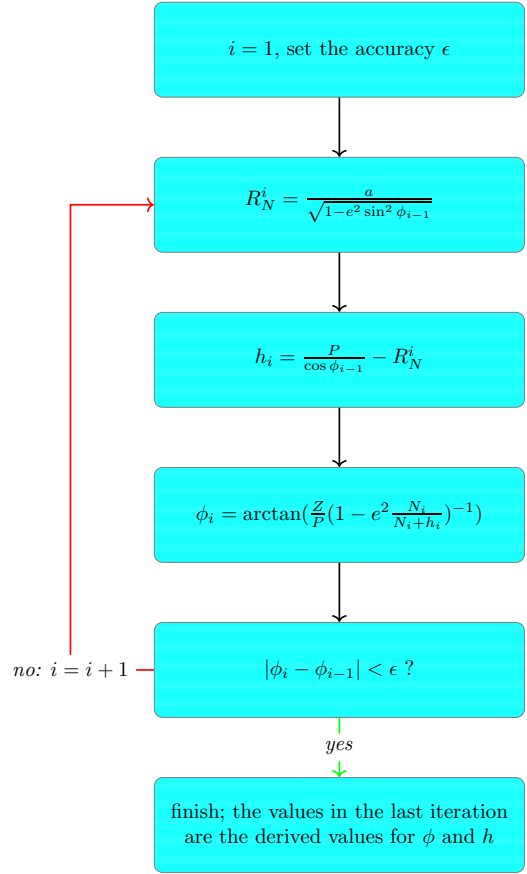
\begin{center}
	\resizebox{0.5\linewidth}{!}{%
		\begin{tikzpicture}[
		node distance = 10mm and 0mm,
		start chain = going below,
		box/.style = {rectangle, rounded corners, draw=gray,
			minimum height=16mm, text width=60mm, align=flush center,
			top color=#1!90, bottom color=#1!90,
			, on chain},
		down arrow/.style = {
			single arrow, draw,
			minimum height=2.5em,
			transform shape,
			rotate=-90,
		}
		]
		\node (n1) [box=cyan]{$i=1$, set the accuracy $\epsilon$};
		\node (n2) [box=cyan]{$R_N^i=\frac{a}{\sqrt{1-e^2\sin^2\phi_{i-1}}}$};
		\node (n3) [box=cyan]{$h_i=\frac{P}{\cos\phi_{i-1}}-R_N^i$};
		\node (n4) [box=cyan]{$\phi_{i}=\arctan(\frac{Z}{P}(1-e^2\frac{N_i}{N_i+h_i})^{-1})$};
		\node (n5) [box=cyan]{$|\phi_{i}-\phi_{i-1}|<\epsilon$ ?};
		\node (n6) [box=cyan]{finish; the values in the last iteration are the derived values for $\phi$ and $h$};
		\draw [black, thick, ->] (n1) edge (n2) (n2) edge (n3)(n3) edge (n4)(n4) edge (n5);
		\path (n5) -- node (yes) {\emph{yes}} (n6);
		\draw[green, thick, ->] (n5) -- (yes) -- (n6);
		
		\path (n5.west) -- ++(-40pt,0pt) node (no) {\emph{no: $i=i+1$}} (n2.west);
		\draw[red, thick, ->] (n5) -- (no) |-(n2);
		\end{tikzpicture}
	}
\end{center}
\end{stp}
\begin{stp}\label{stp2}
The $(X,Y,Z)$ coordinates are transferred to the local geodetic coordinate system, $(X_{LG},Y_{LG},Z_{LG})$, using the derived $\phi$ and $\lambda$ in the previous step in the following relation \cite{Jekeli}, \cite{Krakiwsky}
\begin{equation}\label{eqn8}
\begin{bmatrix}
X_{LG}\\
Y_{LG}\\
Z_{LG}\\
\end{bmatrix}
=S_2~R_2(\phi-\frac{\pi}{2})~R_3(\lambda-\pi)
\begin{bmatrix}
X\\
Y\\
Z\\
\end{bmatrix}
,
\end{equation}
where the $S_2$, $R_2$, and $R_3$ matrices are as the following
\begin{equation}\label{eqn9}
S_2=
\begin{bmatrix}
1 & 0 & 0\\
0 & -1& 0\\
0 & 0 & 1\\
\end{bmatrix}
\end{equation}
\begin{equation}\label{eqn10}
R_2(\theta\in[-\pi,\pi])=
\begin{bmatrix}
\cos(\theta) & 0 & -\sin(\theta)\\
0 & 1 & 0\\
\sin(\theta) & 0 & \cos(\theta)\\
\end{bmatrix}
\end{equation}
\begin{equation}\label{eqn11}
R_3(\theta\in[-2\pi,2\pi])=
\begin{bmatrix}
\cos(\theta) & \sin(\theta) & 0\\
-\sin(\theta) & \cos(\theta) & 0\\
0 & 0 & 1\\
\end{bmatrix}
\end{equation}
\end{stp}
\begin{stp}\label{stp3}
The absolute vertical deflection components, $\xi$ and $\eta$, are calculated as follows \cite{Moritz}
\begin{equation}\label{eqn12}
\begin{split}
\xi&=\frac{1}{R_M\gamma_0}\frac{\partial T}{\partial \phi},\\
\eta&=\frac{1}{R_N\cos\phi\gamma_0}\frac{\partial T}{\partial \lambda},
\end{split}
\end{equation}
where $R_M=\frac{a(1-e^2)}{(1-e^2\sin^2\phi)^{\frac{3}{2}}}$, $R_N=\frac{a}{\sqrt{1-e^2\sin^2\phi}}$, and $\gamma_0$ is the value of gravitational acceleration at the surface of the reference ellipsoid. In addition, $T$ is the residual potential, the value of which can be derived with high precision from the satellite geopotential models, as the following, in which it is calculated up to the degree and order 360
\begin{equation}\label{eqn13}
T(\phi,\lambda,u)=\sum_{n=2}^{360}\sum_{m=0}^{n}\frac{Q_{nm}(i\frac{u}{E})}{Q_nm{i\frac{b}{E}}}(C_{nm}\cos(m\lambda)+S_{nm}\cos(m\lambda))P_{nm}(\sin\phi),
\end{equation}
where $E=a~e$ is the linear eccentricity of the reference ellipsoid, $u=\frac{Z\sqrt{1-e^2\sin^2\phi}}{\sqrt{1-e^2}\sin\phi}$ the third coordinate of the ellipsoid, $C_{nm}$ and $S_{nm}$ respectively the geopotential coefficients, and $P_{nm}$ and $Q_{nm}$ the first and second type of the Legendre functions of degree $n$ and order $m$, respectively.
\begin{rem}\label{rem2}
The value of $\gamma_0$ in \eqref{eqn12} is derived based on the norm of the ellipsoidal gradient of the potential, as follows \cite{Moritz}
\begin{equation}\label{eqn14}
\begin{split}
W(\phi,\lambda,u)&=\sum_{n=0}^{360}\sum_{m=0}^{n}\frac{Q_{nm}(i\frac{u}{E})}{Q_nm{i\frac{b}{E}}}(C_{nm}\cos(m\lambda)+S_{nm}\cos(m\lambda))P_{nm}(\sin\phi),\\
\gamma_0&=||\nabla W||.
\end{split}
\end{equation}
\end{rem}  
\end{stp}
\begin{stp}\label{stp4}
The points in the local geodetic system are transferred to the local astronomical coordinate system \cite{Jekeli}, \cite{Deakin}, \cite{Krakiwsky}. This is done to achieve a more realistic view of the land subsidence or upheave, since the latter coordinate system is directly based on the physics of the earth. To this end, the following transformation is applied to the points in local geodetic coordinate system in \eqref{eqn8}

\end{stp}
\begin{equation}\label{eqn15}
\begin{bmatrix}
X_{LA}\\
Y_{LA}\\
Z_{LA}\\
\end{bmatrix}
=R_1(-\eta)R_2(\xi)~R_3(-\Delta A)
\begin{bmatrix}
X_{LG}\\
Y_{LG}\\
Z_{LG}\\
\end{bmatrix}
,
\end{equation}
in which $\Delta A=\eta\tan\phi$, $R_2$ and $R_3$ matrices with the similar relations as in \eqref{eqn8}, and $R_1$ as the following
\begin{equation}\label{eqn16}
R_1(\theta\in[-2\pi,2\pi])=
\begin{bmatrix}
1 & 0 & 0\\
0 & \cos\theta & \sin\theta\\
0 & -\sin\theta & \cos\theta\\
\end{bmatrix}
\end{equation}
\subsection{Prediction phase and accuracy assessment}
After deriving coordinates in the local astronomical system, the conventional machine learning algorithms can be applied to these coordinates to predict the next outcomes of the time series. As in the previous section, $n$ represents the number of training data and the following dynamic system for the prediction is taken into account (for the meaning of dynamic systems refer to \cite{Ahmed} and \cite{Makridakis})
\begin{equation}\label{eqn17}
Z_{LA}^{i}=f(Z_{LA}^{i-n},Z_{LA}^{i-n+1},...,Z_{LA}^{i-1}),
\end{equation}
where $Z_{LA}^i,~i=n+1,n+2,...$ represents the $i~$th $Z$ coordinate, and $f$ is the function based on which the machine learning algorithm works. In this paper, we use eight different machine learning algorithms, namely, Multi-Layer Perceptron (MLP), Bayesian Neural Network (BNN), Radial Basis Functions (RBF), Gaussian Processes (GP), K-Nearest Neighbor (KNN), Generalized Regression Neural Network (GRNN), Classification And Regression Trees (CART), and Support Vector Regression (SVR). For more information regarding these methods refer to \cite{Alpaydin}, \cite{Watson}, \cite{Awad}, \cite{Nadaraya}, \cite{Orr}, and \cite{Hayes}.

Note that it is important to use reliable indices of the prediction performance. For this reason, three different measures are used, namely, Mean Absolute Scaled Error (MASE), Mean of Absolute Errors (MAE), and Root of Mean Squared Errors (RMSE), defined as the following \cite{Ahmed}, \cite{Makridakis}
\begin{equation}\label{eqn18}
\begin{split}
MASE&=\frac{n-1}{q}\frac{\sum_{i=n+1}^{Q}|Z_{LA}^i-\hat{Z}_{LA}^i|}{\sum_{i=2}^{Q}|Z_{LA}^i-Z_{LA}^{i-1}|},\\
MAE&=\frac{1}{q}\sum_{i=n+1}^{Q}|Z_{LA}^i-\hat{Z}_{LA}^i|,\\
RMSE&=\sqrt{\frac{1}{q}\sum_{i=n+1}^{Q}(Z_{LA}^i-\hat{Z}_{LA}^i)^2},
\end{split}
\end{equation} 
where $q$ and $Q$ denote the total number of predictions and points, respectively, and $Z_{LA}$ and $\hat{Z}_{LA}$ represent the actual and predicted values, respectively. Using these criteria, the prediction accuracy can be assessed. 

The following diagram captures the steps in the proposed algorithm.
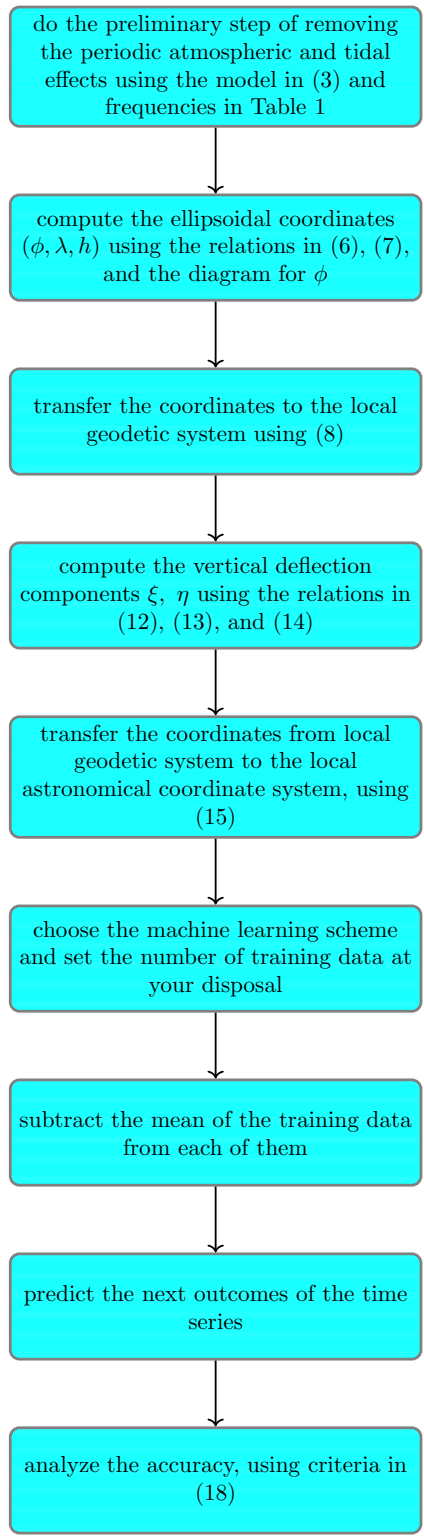
\begin{center}
	\resizebox{0.4\linewidth}{!}{%
		\begin{tikzpicture}[
		node distance = 10mm and 0mm,
		start chain = going below,
		box/.style = {rectangle, rounded corners, draw=gray, very thick,
			minimum height=16mm, text width=60mm, align=flush center,
			top color=#1!90, bottom color=#1!90,
			 on chain},
		down arrow/.style = {
			single arrow, draw,
			minimum height=2.5em,
			transform shape,
			rotate=-90,
		}
		]
		\node (n1) [box=cyan]{do the preliminary step of removing the periodic atmospheric and tidal effects using the model in \eqref{eqn2} and frequencies in Table \ref{tab1}};
		\node (n2) [box=cyan]{compute the ellipsoidal coordinates $(\phi,\lambda,h)$ using the relations in \eqref{eqn6}, \eqref{eqn7}, and the diagram for $\phi$};
		\node (n3) [box=cyan]{transfer the coordinates to the local geodetic system using \eqref{eqn8}};
		\node (n4) [box=cyan]{compute the vertical deflection components $\xi,~\eta$ using the relations in \eqref{eqn12}, \eqref{eqn13}, and \eqref{eqn14}};
		\node (n5) [box=cyan]{transfer the coordinates from local geodetic system to the local astronomical coordinate system, using \eqref{eqn15}};
		\node (n6) [box=cyan]{choose the machine learning scheme and set the number of training data at your disposal};
		\node (n7) [box=cyan]{subtract the mean of the training data from each of them};
		\node (n8) [box=cyan]{predict the next outcomes of the time series};
		\node (n9) [box=cyan]{analyze the accuracy, using criteria in \eqref{eqn18}};
		\draw [black, thick, ->] (n1) edge (n2)(n2) edge (n3) (n3) edge (n4)(n4) edge (n5)(n5) edge (n6)(n6) edge (n7)(n7) edge (n8)(n8) edge (n9);
		
		\end{tikzpicture}
	}
\end{center} 
\section{A comparative analysis of different machine learning algorithms for the prediction of land subsidence or upheave in the proposed algorithm}
The purpose of this section is to represent real examples of the application of the proposed algorithm. For this reason, the mentioned machine learning algorithms in the previous section are used for 14 different time series in Europe. The choice of these time series is influenced by the following factors:

$\bullet$ Inclusion of both continuous and discontinuous time series, to see the effect of gaps on the accuracy assessment.

$\bullet$ Inclusion of both long and short time series.

$\bullet$ Use of time series in different atmospheric and tidal conditions.

We use the same choice of stations in \cite{Kiani1}, which are taken from \cite{Blewitt}. Hence, based on the points mentioned above, the following table represents the result of applying the algorithm (and its aiding machine learning algorithms) to the permanent GNSS stations mentioned in . 
\begin{table}[H]
	\centering
	\caption{The prediction accuracy assessment of eight different machine learning schemes}
	\label{tab2}
	\begin{tabular}{|c|c|c|c|c|c|c|c|c|}\hline
		\cline{1-9} performance criterion & BNN & CART & GP & GRNN & KNN & MLP & RBF & SVR \\
		
		\cline{1-9} $\max(MASE)$ & 0.4027 & 0.2756 & 0.0719 & 0.1610 & 0.1610 & 0.3661 & 24277.6409 & 0.1616 \\
		
		\cline{1-9} $\min(MASE)$ & 0.0066 & 0.0069 & 0.0058 & 0.0067 & 0.0067 & 0.0072 & 1.9072 & 0.0067 \\
		
		\cline{1-9} $mean(MASE)$ & 0.1151 & 0.0610 & 0.0229 & 0.0476 & 0.0477 & 0.0696 & 1848.7074 & 0.0477 \\
		
		\cline{1-9} $\max(MAE)$ & 0.2559 & 0.1752 & 0.0427 & 0.1024 & 0.1024 & 0.2327 & 15432.6477 & 0.1027  \\
		
		\cline{1-9} $\min(MAE)$ & 0.0041 & 0.0043 & 0.0039 & 0.0042 & 0.0042 & 0.0044 & 0.2152 & 0.0042\\
		
		\cline{1-9} $mean(MAE)$ & 0.0591 & 0.0303 & 0.0102 & 0.0232 & 0.0232 & 0.0332 & 1183.2102 & 0.0233 \\
		
		\cline{1-9} $\max(RMSE)$ & 0.2571 & 0.1769 & 0.1399 & 0.1399 & 0.1399 & 0.2340 & 19899.7476 & 0.1399 \\
		
		\cline{1-9} $\min(RMSE)$ & 0.0053 & 0.0055 & 0.0049 & 0.0054 & 0.0054 & 0.0055 & 0.2659 & 0.0054 \\
		
		\cline{1-9} $mean(RMSE)$ & 0.0650 & 0.0423 & 0.0221 & 0.0353 & 0.0353 & 0.0445 & 1525.0325 & 0.0354\\
		
		\hline
	\end{tabular}
\end{table}
\begin{table}[H]
	\centering
	\caption{The permanent GNSS stations and they subsidence or upheave prediction status}
	\label{tab4}
	\begin{tabular}{|c|c|c|c|}\hline
		\cline{1-4} permanent GNSS station & time span & data continuity & state \\
		
		\cline{1-4} A Coruna (Spain) & 1998-2020 & no & subsidence\\
	
	\cline{1-4} Ajaccio (France) & 2000-2020 & no & upheave\\
	
	\cline{1-4} Bacau (Romania) & 2006-2020 & yes & subsidence\\
	
	\cline{1-4} Borowa Gora (Poland) & 1996-2020 & yes & subsidence \\
	
	\cline{1-4} Svetloe (Russian Federation) & 1997-2020 & yes & upheave\\
	
	\cline{1-4} Morpeth (United Kingdom) & 1996-2020 & no & upheave\\
	
	\cline{1-4} Kunzak (Czech Republic) & 2005-2020 & yes & subsidence\\
	
	\cline{1-4} Maartsbo (Sweden) & 1996-2020 & yes & upheave\\
	
	\cline{1-4} Mariupol (Ukraine) & 2013-2020 & yes & subsidence\\
	
	\cline{1-4} Matera (Italy) & 1994-2020 & yes & subsidence\\
	
	\cline{1-4} Kirkkonummi (Finland) & 2013-2020 & yes & upheave\\
	
	\cline{1-4} Modra-Piesok (Slovak Republic) & 2007-2020 & yes & subsidence \\
	
	\cline{1-4} Nicosia (Cyprus) & 1997-2020 & no & upheave\\
	
	\cline{1-4} Athens (Greece) & 2006-2020 & yes & upheave\\
		
		\hline
	\end{tabular}
\end{table}
It is important to note that the least levels of accuracies are for the discontinuous data. It is because of them that the overall accuracy is in the centimeter level. If only continuous data were available, the accuarcy level would be in the millimeter level. 
\subsection{Comparison with the Theta statistical method}
We perform the same analyses with the traditional statistical method called Theta \cite{Assimakopoulos}, to evaluate the relative accuracy of the machine learning algorithms against the statistical methods. After performing the assessments, we get the following results.
\begin{table}[H]
	\centering
	\caption{The prediction accuracy assessment of eight different machine learning schemes}
	\label{tab3}
	\begin{tabular}{|c|c|}\hline
		\cline{1-2} performance criterion & Theta \\
		
		\cline{1-2} $\max(MASE)$ & 119213853.3002\\
		
		\cline{1-2} $\min(MASE)$ & 62.5534 \\
		
		\cline{1-2} $mean(MASE)$ & 72.2209\\
		
		\cline{1-2} $\max(MAE)$ & 1647636.6561\\
		
		\cline{1-2} $\min(MAE)$ & 2.2602\\
		
		\cline{1-2} $mean(MAE)$ & 2.6114\\
		
		\cline{1-2} $\max(RMSE)$ & 33448426.7666\\
		
		\cline{1-2} $\min(RMSE)$ & 20.7481\\
		
		\cline{1-2} $mean(RMSE)$ & 23.9242\\
		
		\hline
	\end{tabular}
\end{table}
As it can be understood from the Tables \ref{tab2} and \ref{tab3}, Theta method is much less accurate than the machine learning methods. The GP algorithm is the most accurate machine learning method, since it has the lowest values of $mean(RMSE)$ and $MASE$. On the other hand, as \cite{Kiani1} also asserts, the RBF algorithm is the least accurate method. Once again it is confirmed that the RBF algorithm is not suitable for the GNSS position time series prediction. Other methods have better performances than RBF. But they are less accurate than GP. Overall, it can be said that the GP is the best choice for the prediction of GNSS position time series.
\section{Conclusion}
An algorithm is presented to infer the land subsidence or upheave from GNSS position time series data. After initial preprocessing, which include the removal of atmospheric and tidal effects, the established machine learning algorithms are used to predict the next outcomes of the time series. It is shown that the algorithm can achieve up to the millimeter accuracy, in a study for 14 different time series. A comparison between the machine learning algorithms and the statistical Theta method for prediction reveals that the machine learning algorithms are much more accurate. 

The results of this paper are important in that they present us with a powerful method that can be used alongside other methods of subsidence or upheave prediction in an area. This would open up a new phase of research for geodetic community and geoscientists who work in this area.    
    
\end{document}